\documentclass[aps,prb,twocolumn,superscriptaddress,showpacs,floatfix]{revtex4}
\usepackage{graphicx}
\usepackage{amsmath}
\usepackage{url}

\newcommand{\ep}[0]{\varepsilon_p}
\newcommand{\om}[0]{\omega}
\newcommand{\nag}{\phantom{\dag}}
\newcommand{\las}[0]{\langle}
\newcommand{\ras}[0]{\rangle}

\DeclareMathOperator{\trace}{Tr}

\DeclareMathOperator{\I}{i}
\DeclareMathOperator{\E}{e}
\DeclareMathOperator{\re}{Re}

\graphicspath{{.}{../}{./eps/}}

\begin{document}
\title{Optical absorption and activated transport in polaronic systems} 
\author{G. Schubert}
\affiliation{Institut f\"ur Physik, Ernst Moritz Arndt Universit\"at
  Greifswald, 17487 Greifswald, Germany}
\author{G. Wellein}
\affiliation{Regionales Rechenzentrum Erlangen, 
  Universit\"at Erlangen, 91058 Erlangen, Germany}
\author{A. Wei{\ss}e}
\affiliation{School of Physics, The University of New South Wales,
  Sydney, NSW 2052, Australia}
\author{A. Alvermann}
\affiliation{Institut f\"ur Physik, Ernst Moritz Arndt Universit\"at
  Greifswald, 17487 Greifswald, Germany}
\author{H. Fehske}
\affiliation{Institut f\"ur Physik, Ernst Moritz Arndt Universit\"at
  Greifswald, 17487 Greifswald, Germany}
\date{\today}

\begin{abstract}
  We present exact results for the optical response in the
  one-dimensional Holstein model. In particular, by means of a refined
  kernel polynomial method, we calculate the ac and dc electrical
  conductivities at finite temperatures for a wide parameter range of
  electron phonon interaction.  We analyze the deviations from
  the results of standard small polaron theory in the intermediate
  coupling regime and discuss non-adiabaticity effects in detail.
\end{abstract}
%71.38.-k Polarons and electron-phonon interactions 
%71.38.Ht Self-trapped or small polarons
%74.25.Fy Transport properties (electric and thermal conductivity, thermoelectric effects, etc.)

\maketitle
\section{Introduction}
The investigation of transport properties has been playing a central
role in condensed matter physics for a long time.  In recent years
optical spectroscopy, for example, has contributed a lot to unravel
the complex physics of highly correlated many-body systems, such as
the one-dimensional (1d) MX chains,\cite{BS98} the quasi-2d
high-temperature superconducting cuprates,\cite{Da94} or the 3d
colossal magneto-resistive manganites.\cite{JTMFRC94} That way, optical
measurements proved the importance of electron-phonon (EP)
interactions in all these materials and, in particular, corroborated
polaronic scenarios for modeling their electronic transport properties
at least at high temperatures.\cite{Emi93,AM95,WMG98}

Polarons are quasi-particles composed of an electron and the
surrounding ions which in a polar solid, provided the electron lattice
interaction is sufficiently strong, are displaced from their
equilibrium positions due to the presence of the electron. This
bootstrap relation between electron and lattice displacement makes the
particle heavy, because it has to drag with it the potential well of
the phonons.  Polaron motion is largely understood and has been worked
out theoretically in two important limits:  In the first case the
electronic bandwidth is large and there is a only a slight change in
the particle's effective mass due to the EP coupling. These
quasi-particles are called large polarons or Fr\"ohlich polarons. In
the second case it is assumed that the bandwidth is small, whereas the
EP interaction is strong and short-ranged. Now polaronic effects trap
the electron at a certain lattice site and the size of the
quasi-particle becomes comparable to the inter-atomic lattice spacing.
Thermally activated hopping will necessarily be the dominant transport
process of such small or Holstein-type polarons.  Although there are
experimental systems with clear large and small polaron
characteristics, most of the above-mentioned novel materials belong to
the transition region between these two limiting cases.  Here the
relevant energy scales are not well-separated and perturbative
approaches cannot describe the complicated transport mechanisms
adequately.

A recent non-perturbative dynamical mean-field study of the Holstein
model in infinite dimensions\cite{FC03} reports quantitative
discrepancies of the temperature dependence of the resistivity from
standard polaron formulas, but the dynamical mean-field approach
inherently does not account for vertex corrections to the
conductivity and for longer ranged hopping processes induced by the EP
interaction. On the other hand, exact numerical investigations of the
Holstein model provided quite a number of reliable results for the
zero-temperature optical conductivity in the 1d and 2d (extended)
Holstein models.\cite{FLW97,ZJW99}

Motivated by this situation, in the present work we use a recently
developed extension of the Kernel Polynomial Method
(KPM),\cite{SRVK96,WWAF05} a refined numerical Chebyshev expansion
technique, to compute both the dc and ac hopping conductivities at
finite temperatures without any serious approximation. 
\section{Model and Method}
Our starting point is the 1d tight-binding 
Holstein Hamiltonian,\cite{Ho59a}
\begin{equation}
  H = -t \sum_{\las i,j\ras} c^\dag_i c^{\nag}_j 
  -\sqrt{\ep\om_0} \sum_i (b^\dag_i + b^{\nag}_i) n_i^{\nag} 
  +\hbar\om_0\sum_i b^\dag_i b^{\nag}_i\,,
\label{homo}
\end{equation}
describing a single electron coupled locally to a dispersion-less
optical phonon mode, where $c^\dag_i$ ($b^\dag_i$) denotes the
corresponding fermionic (bosonic) creation operator, and $n_i=c^\dag_i
c^{\nag}_i$. Setting $\hbar=1$ and measuring all energies in units of
the nearest-neighbor hopping integral $t$, the physics of the model is
determined by the dimensionless EP coupling constants
\begin{equation}
  \lambda=\ep/2t\;\;\; \mbox{and}\;\;\;g^2=\ep/\om_0 
  \label{coco}
\end{equation}
in the adiabatic ($\om_0/t \ll 1$) and anti-adiabatic ($\om_0/t \gg
1$) regimes, respectively. In 1d the crossover from large to small
polaron behavior takes place at $\lambda \simeq 1$ ($g^2 \simeq 1$) in
the former (latter) case.\cite{WF97,CSG97} Whether large polarons form
in the Holstein model for $\text{d}>1$ is still under
debate.\cite{Emi86,FRWM95,CFI99}

Addressing the linear response of our system to an external 
(longitudinal) electric field we consider  
the Kubo formula for the electrical conductivity
at finite temperatures, which is\cite{Mah00}
\begin{equation}
  \re \sigma(\omega) = \pi
  \sum_{m,n}^{\infty} \frac{\E^{-\beta E_n} - \E^{-\beta E_m}}{Z L \omega} 
  \,|\langle n|\hat{\jmath}|m\rangle|^2 \,
  \delta(\omega - \omega_{mn})\,.
  \label{si_1}
\end{equation}
Here $Z = \sum_{n}^{\infty} \E^{-\beta E_n}$ is the partition function
and $\beta=T^{-1}$ denotes the inverse temperature.  Since the
Holstein Hamiltonian~(\ref{homo}) involves bosonic degrees of freedom,
the Hilbert space even of a finite $L$-site system has infinite
dimension. In practice, however, the contribution of highly excited
phonon states is negligible at the relevant temperatures, and the system
is well approximated by a truncated phonon space with at most
$M(\lambda,g,\omega_0;T)$ phonons.\cite{BWF98} Then
$|n\rangle$ and $|m\rangle$ are the eigenstates of $H$ within our
truncated $D$-dimensional Hilbert space, $E_n$ and $E_m$ are the corresponding
eigenvalues, and $\omega_{mn} = E_m - E_n$. In Eq.~(\ref{si_1}) the
current operator has the standard hopping form,
%\begin{equation}
$\hat{\jmath}=\I\, e  t \sum_i (c^\dag_i c^{\nag}_{i+1} - c^{\dag}_{i+1} c^{\nag}_i)\,,$
%\label{j}
%\end{equation}
and connects states with different parity.\cite{Ba02}  
Thus, assuming that the ground state is non-degenerate, 
the expectation value $\langle
0|\hat{\jmath}|0\rangle$ vanishes in the absence of an external electric
field.  The limit $\om\to 0$ of (\ref{si_1}) yields the dc
conductivity, whereas the optical absorption is given by the finite
frequency data. For small polarons both results are interesting and
have been evaluated analytically at an early stage.\cite{LF62} At
$T=0$, the regular part of the optical conductivity,
$\sigma^{\text{reg}}(\omega) = \frac{\pi}{L} \sum_{n>0}
\omega_{n0}^{-1} |\langle n|\hat{\jmath}|0\rangle|^2 \delta(\omega -
\omega_{n0})$, was calculated for finite 1d and 2d lattices with
periodic boundary conditions (PBC) in a wide parameter range of
the Holstein model, using a combination of the Lanczos algorithm and
the KPM.\cite{FLW97} 

\begin{figure}[t]
  \begin{center}
   \includegraphics[width=\linewidth,bb=200 90 885 370,clip]{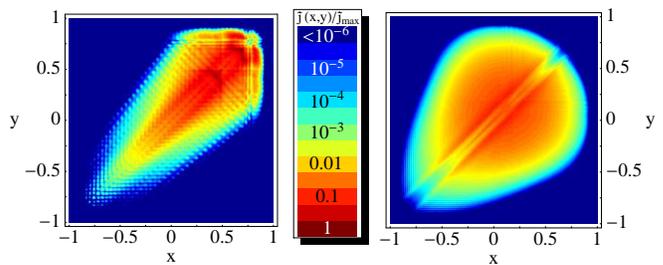}
  \end{center}
  \caption{(Color online) Renormalized current operator density  
    $\tilde{\jmath}(x,y)$ used in the 2d KPM. 
    Data obtained for the 1d Holstein model with
    $\lambda=0.2$ (left panel) and $\lambda=2.0$ (right panel) at
    $\om_0/t=0.4$ ($L=6$, $M=50$; PBC).}\label{f_od}
\end{figure}

At finite temperatures, a similar straight-forward expansion of the
conductivity is spoiled by the presence of the Boltzmann factors and
the contribution of all matrix elements between eigenstates of the
system. Instead, it turns out that a new 
generalised KPM scheme~\cite{WWAF05,We04} can be
based upon a current operator density
\begin{equation}
 j(x,y) = \sum_{m,n} |\langle n|\hat{\jmath}|m\rangle|^2 
  \ \delta(x-E_n)\ \delta(y-E_m)\,.
\label{j_xy}
\end{equation}
Being a function of two variables, $j(x,y)$ can be expanded by a
two-dimensional KPM,
\begin{equation}
  \tilde{\jmath}(x,y) = \sum\limits_{k,l=0}^{N-1}\frac{
    \mu_{kl} w_{kl} g_k^J g_l^J T_k(x) T_l(y)
  }{\pi^2 \sqrt{(1-x^2)(1-y^2)}}\,,
\label{jt_xy}
\end{equation}
where the tilde refers to a rescaling of energy ($H\to\tilde{H}$) 
that maps the spectrum of $H$ into the domain~$[-1,1]$ of the
Chebyshev polynomials of first kind $T_k(x)$.  
The finite order~$N$ of the expansion
leads to Gibbs oscillations which can be damped by introducing
appropriate damping factors. Here we use $g_n^J$ derived from the
Jackson kernel\cite{Ja12,SRVK96}, and $N=512$ throughout the paper. 
The core of the numerical work is
the iterative calculation of the moments
$\mu_{kl} = \trace(T_k(\tilde H) \hat{\jmath} T_l(\tilde H) \hat{\jmath})$,
where the trace can be replaced by an average over a relatively small
number of random vectors $|r\rangle$.\cite{DS93} Finally, the factors
$1/w_{kl} = (2-\delta_{k0})(2-\delta_{l0})$ account for the correct
normalisation.  Given the operator density $j(x,y)$ we find the
optical conductivity by integration
\begin{equation}
    \re\sigma(\omega) \!=  \!
    \frac{\pi}{ZL\omega} \int\limits_{-\infty}^{\infty}
    j(y+\omega,y)
    \big[\E^{-\beta y} - \E^{-\beta (y+\omega)}\big]\, dy\,.
\label{si_2}
\end{equation}
The partition function $Z=\int_{-\infty}^{\infty} \rho(E)\exp (-\beta
E)$ is easily obtained by integrating over the density of states
$\rho(E) = \sum_{n=0}^{D-1} \delta (E-E_n)$, which can be expanded in
parallel to $\tilde{\jmath}(x,y)$. Note the main advantage of this
approach: The current operator density that enters the conductivity is
the same for all temperatures, i.e., it needs to be expanded only
once. Figure~\ref{f_od} exemplifies the different form of
$\tilde{\jmath}(x,y)$ in the weak and strong EP coupling regimes.

\begin{figure}[b]
  \centering 
  \includegraphics[width=0.98\linewidth,clip]{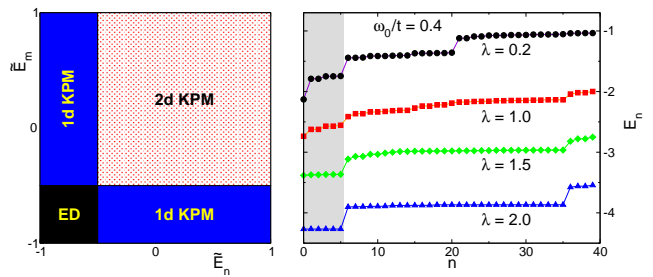}
  \caption{(Color online) Schematic setup for the calculation
    of the finite-temperature optical conductivity (left panel).
    Lowest eigenvalues of the Holstein Hamiltonian for $L=6$, $M=25$, and PBC
    (right panel).
    The shaded area marks the six lowest eigenvalues to be separated from
    the rest of the spectrum.}\label{f_setup}
\end{figure}

At very low temperatures, the numerical evaluation of
expression~(\ref{si_2}) requires some caution, since the Boltzmann
factors heavily amplify small numerical errors in $j(y+\om,y)$. We can
avoid these problems, occurring mainly at the lower bound of the spectrum,
by treating the contributions of the ground state and some of the
lowest excitations separately. This is illustrated
in Fig.~\ref{f_setup}. We split the optical conductivity into three
parts,
\begin{equation} 
  \re\sigma(\omega) = 
  \re\sigma^{\text{ED}}(\omega) + 
  \re\sigma^{\text{1d}}_{\text{KPM}}(\omega) + 
  \re\sigma^{\text{2d}}_{\text{KPM}}(\omega)\,,
\label{si_3}
\end{equation}
where the first contribution describes the transitions (matrix
elements) between the $S$ separated eigenstates, the second part those
between the separated states and the rest of the spectrum, which can
be expressed as standard 1d~KPM expansions, and finally the transitions within
the remaining $D-S$ states of the spectrum, handled by a 2d expansion.
Using the projection operator $P = 1 - \sum_{s=0}^{S-1}
|s\rangle\langle s|$, the moments for the contributions
$\re\sigma_{\text{1d}}(\omega)$ read $\mu_k^n = \langle
n|\hat{\jmath}PT_k(\tilde H)P\hat{\jmath}|n\rangle$.  Of course, the number of
states one has to separate depends on the physical situation.  The
right panel of Fig.~\ref{f_setup} gives the lowest eigenvalues of the
Holstein model at various coupling strengths. In the strong-coupling
regime ($\lambda \gg 1$) states belonging to the lowest small polaron
band have almost the same energy as the ground state and therefore
should be treated separately (cf.  the curves for $\lambda=1.5$
and~2). Obviously the situation is far less dramatic at weak EP
couplings.

\begin{figure}[bt]
  \centering 
  \includegraphics[width=0.9\linewidth,clip]{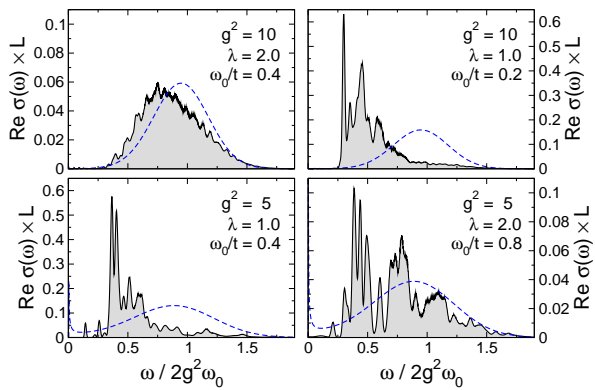}
   \caption{(Color online) 
     Optical conductivity in the 1d Holstein model at $T=0$ 
     (in units of $\pi e^2 t^2$) 
     compared to the analytical small polaron result 
     Eq.~(\ref{aspt}) [dashed blue lines].
     Exact diagonalization data (ED) are obtained for a system with $L=6$
     and $M=45$;  
     $\sigma_0$ is determined to give the same
     integrated spectral weight of the $\omega>0$ (regular part)   
     of $\re \sigma$.}
   \label{acsigma_kpm_reik}
\end{figure}
\section{Numerical results and discussion}
\subsection{ac concductivity}
We now apply our numerical scheme to the calculation of the optical
absorption in the 1d Holstein model. 
The results for $\re \sigma(\om)$ 
and possible deviations from established polaron theory
are important for relating theory and experiment.
The standard description of small polaron transport\cite{RH67,Emi93} 
provides the ac conductivity at $T=0$ as 
\begin{equation}
  \re \sigma (\omega) = 
  \frac{\sigma_0}{\sqrt{\ep\omega_0}} \frac{1}{\omega} 
  \exp \left[- \frac{(\omega-2 \ep)^2}{4 \ep \omega_0}\right]
  \; .
\label{aspt}
\end{equation}
For sufficiently strong coupling this formula predicts a weakly asymmetric
Gaussian absorption peak centered at $\omega = 2 \ep$.
A similiar analytical formula can be derived for finite 
temperatures.\cite{BB85,Mah00}

Starting at {\it zero temperature}, Fig.~\ref{acsigma_kpm_reik} 
shows $\re \sigma(\omega)$ for various EP-parameters.
For $\lambda=2$ and $\om_0/t=0.4$, i.e., at rather large EP coupling,
but not in the extreme small polaron limit, we find 
also a pronounced maximum in the low-temperature optical
response, which, however, is located, 
somewhat below $2\ep=2g^2\om_0$, being the value for small
polarons at $T=0$. At the same time, the line-shape is more asymmetric
than in standard polaron theory, with a weaker decay at the
high-energy side, which fits even better the experimental behavior
observed in polaronic materials such as TiO$_2$.\cite{KMF69}
Varying the parameters significant discrepancies 
to a Gaussian-like absorption are found.
Then the polaron motion is not adequately described 
as hopping of a self-trapped carrier almost localized on a single
lattice site.

\begin{figure}[bt]
  \centering 
  \includegraphics[width=0.9\linewidth,clip]{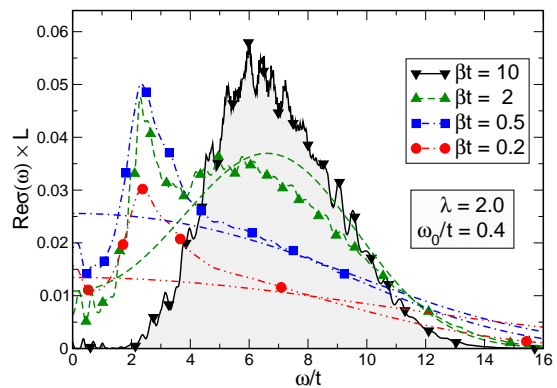}
   \caption{(Color online) Optical absorption by Holstein polarons at finite
  temperatures in the adiabatic regime ($L=6$, $M=45$).
    Dashed curves give the analytical result  
    for finite-temperature small polaron transport.\cite{BB85,Mah00}
    The deviations observed for high excitation energies
    at very large temperatures are caused by the necessary 
    truncation of the phonon Hilbert space in ED.}
\label{acsigma_sc_lf}
\end{figure}

At {\it finite temperature} two different transport mechanism 
can be distinguished. Clearly, 
coherent transport, which for large EP couplings is related to
diagonal (zero-phonon) transitions within the lowest extremely narrow
polaron band, will be negligible at high temperatures. For
instance, the amplitude of the current matrix elements between the
degenerate states with momentum $K=\pm\pi/3$ ($K=0,\, \pm\pi/3,\,
\pm2\pi/3$, and $\pi$ are the allowed wave numbers of a 6-site system
with PBC) is of the order of $10^{-7}$ only.  Whereas phase coherence
is maintained during a diagonal transition, the particle loses its
phase coherence if its motion is triggered by (multi-) phonon
absorption and emission processes. These so-called non-diagonal
transitions which, of course, can take place also with zero energy
transfer, become more and more important as the temperature increases.
Accordingly the main transport mechanism 
is thermally activated hopping, where
each hop becomes a statistically independent event. In the small
polaron limit, where the polaronic sub-bands are roughly separated by
the bare phonon frequency (cf.  Fig.~\ref{f_setup}, right panel), this
happens for $T\gtrsim \om_0$.  Let us consider the activated regime
in more detail (cf. Fig.~\ref{acsigma_sc_lf}: 
With increasing temperatures we observe a substantial
spectral weight transfer to lower frequencies, and an increase of the
zero-energy transition probability in accordance with
previous results.\cite{Na63} 

In addition, we find a strong resonance
in the absorption spectra at about $\om\sim 2t$, which can be easily
understood using a configurational coordinate picture. Placing a
homogeneous lattice distortion $u$ at $L_u$ consecutive sites by
applying the unitary transformation
$S^\dagger(u)=\prod_i^{L_u}S^\dagger_i(u)=\prod_i^{L_u} 
\exp[u(b_i^\dagger -b_i^{\nag})]$, the
transformed Holstein Hamiltonian takes the form $\bar{H}= \langle
0|S^\dagger(u)HS(u)|0\rangle_{\rm ph}= -t \sum_{\las i,j\ras} c^\dag_i
c^{\nag}_j -2\sqrt{\ep\om_0} u \sum_i^{L_u} n_i + \om_0u^2L_u$.  In
the adiabatic strong-EP-coupling regime, the ground state can be
approximated as an electron localized at a certain single site with
the lattice being in a shifted oscillator state ($L_u=1$). That is,
$|\Psi_0\rangle= |1\rangle_{el}\otimes S^{\dag}_1(g) |0\rangle_{ph}$
and $E_0=-\ep$, in accordance with the Lang-Firsov approximation.  Now
let us consider excitations from this ground-state, where the lattice
distortion spreads over two neighboring sites ($L_u=2$) and the
electron is in a symmetric or antisymmetric linear combination of
$|1\rangle_{el}$ and $|2\rangle_{el}$, i.e., the particle is mainly
located at sites 1 and 2 but delocalized between these sites. We then find
$|\Psi_{1,\pm} \rangle= (|1\rangle_{el}\pm |2\rangle_{el})\otimes
S^{\dag}_2(g/2) S^{\dag}_1(g/2) |0\rangle_{ph}$ and $E_{1,\pm} =\mp
t-\ep/2$. Whereas the (potential) energy related to the displacement
field is reduced to $-\ep/L_u$, the kinetic energy comes into play
since hopping processes between 1 and 2 are allowed.  The current
operator $\hat{\jmath}$ connects these different-parity states with perfect
overlap $|\langle \Psi_{1,+}|\hat{\jmath} | \Psi_{1,-}\rangle|=(et)^2$,
giving rise to a strong signal in the optical absorption.  Note that
the excitation energy $\om_{1-,1+}=2t$ is independent of $\ep$. In
order to activate these transitions thermally, the electron has to
overcome the ``adiabatic'' barrier $\Delta=E_{1+}-E_0=\ep/2-t$. A
finite phonon frequency will relax this condition. From
Fig.~\ref{acsigma_sc_lf}, we find the signature to occur above
$T\gtrsim 0.5t$. 
  Obviously this feature is absent in the standard small-polaron transport
  description which essentially treats the polaron as a quasiparticle
  without resolving its internal structure.
  Owing to the infinite number of neighboring sites it is also absent
  in the DMFT calculation. 
Of course, one could also extend this scenario to
excitations where the electron is delocalized over more than two
distorted lattice sites, but 
%presumably 
for the present parameters the
signature of these weakly bound states would be rather small. 
%if they exist at all.

\begin{figure}[t]
  \centering 
  \includegraphics[width=0.9\linewidth,clip]{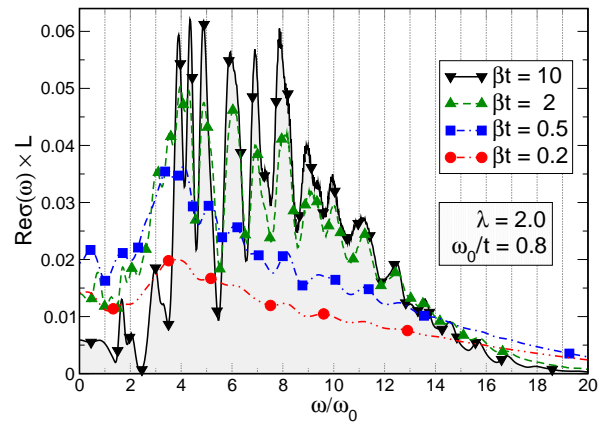}
  \caption{(Color online) Optical absorption by Holstein polarons at
    finite temperatures in the non-adiabatic regime ($L=6$, $M=30$). 
    Note that now the
    abscissa is scaled with respect to the phonon frequency.}
\label{acsigma_sc_hf}
\end{figure}\begin{figure}[b]
  \centering 
  \includegraphics[width=0.9\linewidth,clip]{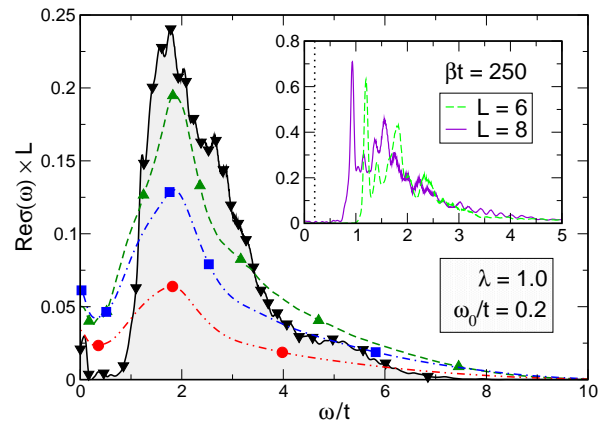}
  \caption{(Color online) Optical absorption in the adiabatic 
    intermediate EP-coupling regime (the notation is the
    same as in Fig.~\ref{acsigma_sc_lf}; again we use
    $L=6$, $M=40$). The inset illustrates
    the finite-size dependence of $\re \sigma(\omega)$ for 
     $T\sim 0$ (the vertical dotted line gives 
    the phonon absorption threshold). It demonstrates that 
    the gap observed at low frequencies and temperatures 
    is clearly a finite-size effect, i.e.,   
    at weak-to-intermediate couplings the discrete
    electronic levels of our finite system show up 
    in the conductivity spectra.  These effects, of course, 
    are of minor importance at larger EP couplings, 
    where the polaronic bandwidth is strongly reduced, as well as for 
    high temperatures.}  
  \label{acsigma_ic_lf}
\end{figure}
Entering the non-adiabatic regime of large phonon frequencies
at fixed $\lambda=2$, the pattern of sub-bands separated
roughly by $\om_0$ becomes more pronounced, but is, of course, washed
out at higher temperatures (see Fig.~\ref{acsigma_sc_hf} for
$\om_0/t=0.8$).  In Fig.~\ref{acsigma_sc_hf} the average number of
phonons contained in the ground state ($\propto g^2$) is smaller
($g^2=5$) than in the previous case where $g^2=10$ ($\om_0/t=0.4$).
This also concerns the activated region $\om_0/t\lesssim
T\lesssim\Delta$ but for these parameters the simple adiabatic
picture anticipated above breaks down anyway.

Now let us decrease the EP coupling strength at small phonon
frequencies $\om_0/t=0.2$ keeping $g^2=10$ fixed.  Results for the
optical response in the vicinity of the large to small polaron
crossover ($\lambda=1$) are depicted in Fig.~\ref{acsigma_ic_lf}.
Here the small polaron maximum has almost disappeared and the
$2t$-absorption feature can be activated at very low temperatures
($\Delta \to 0$ for the two-site model with $\lambda=1$).  The overall
behavior of $\re\sigma(\omega)$ resembles that of polarons of
intermediate size. At high temperatures these polarons will dissociate
readily and the transport properties are equivalent to those of
electrons scattered by thermal phonons.  Let us emphasize that
many-polaron effects become increasingly important in the
large-to-small polaron transition region.\cite{Hoea04p} As a result,
polaron transport might be changed entirely compared to the
one-particle picture discussed so far.
\subsection{Sum rules}
Before we consider the temperature dependence of the dc conductivity, 
it is useful to test the sum rules for the real part of the optical
response.  First we have the so-called $f$-sum rule,
\begin{equation} 
  S_{\rm tot}:=\int_{-\infty}^\infty \re\sigma(\omega) d\om = 
  - \frac{\pi e^2}{L} E_{\rm kin}\,,
  \label{fsum1}
\end{equation}
which relates the $\omega$-integrated $\re\sigma(\omega)$ to the
kinetic energy $E_{\rm kin}=-t\sum_{\langle i,j \rangle} \langle
c_i^{\dag}c_j^{\nag}\rangle_T$.  Note that the $\omega=0$ (Drude)
contribution is included in~(\ref{fsum1}).  The second sum-rule for
$\re\sigma(\omega)$ is
\begin{equation} 
  \int_{0}^\infty \omega \re\sigma(\omega) d\om = 
  \frac{\pi}{L}  \langle \hat{\jmath}^{2}\rangle_T\,. 
  \label{fsum2}
\end{equation}
Throughout our calculations Eq.~(\ref{fsum2}) was fulfilled within
numerical accuracy, where the thermal average $\langle
\hat{\jmath}^{2}\rangle_T$ was determined again using a 1d KPM.
Testing sum rule~(\ref{fsum1}) we gain important
information about finite-size effects.

\begin{figure}[t]
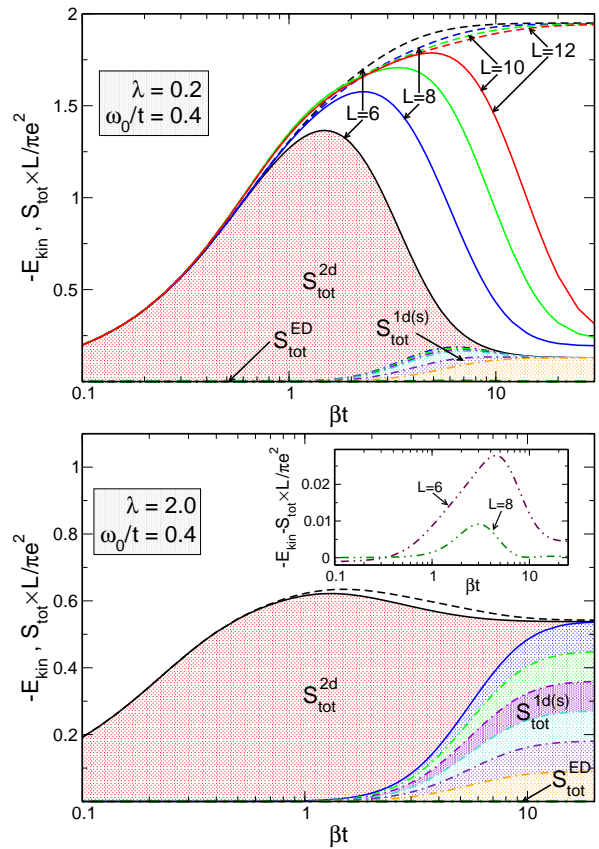

  \centering 
  \includegraphics[width=0.9\linewidth,clip]{fc7a.eps}\\
  \includegraphics[width=0.9\linewidth,clip]{fc7b.eps}
  \caption{(Color online) Conformance of the $f$-sum rule~(\ref{fsum1}) 
    in the weak (upper panel) and strong (lower panel) EP coupling regimes
    of the 1d Holstein model. 
    Results for $S_{\rm tot}$, partial $S_{\rm tot}$'s and $E_{\rm kin}$ are given by solid,
    dot-dashed and dashed lines, respectively. 
    For further explanation see text.}\label{fsum1rule}
\end{figure}

Figure~\ref{fsum1rule} compares $E_{\rm kin}$ and $S_{\rm tot} L/ \pi e^2$
obtained from our finite-cluster calculation.  At weak EP couplings
the kinetic energy is a strictly monotonic increasing function of
temperature and becomes strongly suppressed at high temperatures due
to scattering of the electron by thermal phonons. Whereas the $f$-sum
rule is almost perfectly fulfilled for smaller values of $\beta$, we
found pronounced deviations at low temperatures which, without doubt,
can be assigned to the finite size of our Holstein ring (cf.
the $L$ dependence of $S_{\rm tot}$ shown in the upper panel of
Fig.~\ref{fsum1rule}, and also the discussion of the dc conductivity
below). Clearly finite-size effects become important when the
temperature is comparable to the energy gaps in the spectrum of $H$
(see Fig.~\ref{f_setup}).  At strong EP couplings the transport is
hopping-dominated and the kinetic energy exhibits a maximum at a
finite temperature that can be related to the thermal activation
energy of polarons. Now, in view of the narrow small polaron band (cf.
Fig.~\ref{f_setup}), finite size gaps are small and the $f$-sum rule
is fulfilled down to very low temperatures. Small deviations appear to
vanish rapidly with increasing system size (see inset, lower panel of
Fig.~\ref{fsum1rule}). In order to analyze the contributions of
different transport processes to $E_{\rm kin}$ in some more detail, we
have decomposed $S_{\rm tot}=S_{\rm tot}^{\text{ED}}+\sum_{s=0}^{S-1}
S_{\rm tot}^{\text{1d} (s)} +S_{\rm tot}^{\text{2d}}$ in analogy to
Eq.~(\ref{si_3}).  Figure~\ref{fsum1rule} shows that the coherent
(intra-band) contribution ($\propto S_{\rm tot}^{\text{ED}}$) is almost
negligible at the temperatures considered. Inter-band transitions
connecting eigenstates of the lowest polaron band to higher excited
states (1d KPM) are the determining factor at low $T$. If the
renormalised polaron bandwidth is small enough, all states in the band
are equally populated, leading to pretty much the same values of
$S_{\rm tot}^{\text{1d} (s)}$.  At high temperatures, of course, the
transitions covered by the 2d KPM are predominant.

\begin{figure}[t]
  % \centering 
  \hspace*{0.1cm}
  \includegraphics[width=0.93\linewidth,clip]{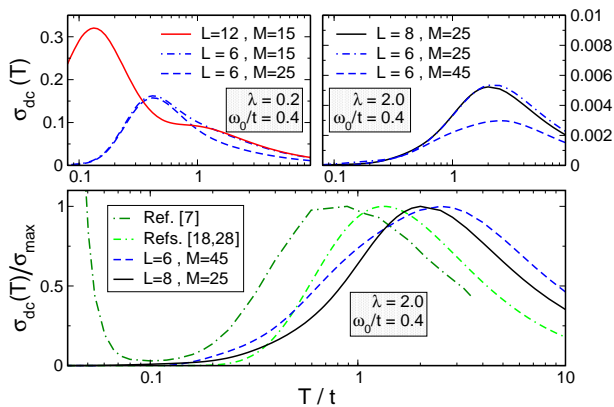}\\[0.1cm]
  \caption{(Color online) Top row: DC conductivity as a function of
    temperature  at weak (left panel) 
    and strong (right panel) EP coupling.
    The bottom panel shows, for strong coupling, a comparison 
    with recent DMFT results\cite{FC03} and 
    standard polaron theory\cite{Mah00}. }\label{sigma_dc}
\end{figure}
\subsection{dc conductivity}
The dc conductivity is obtained by taking the limit $\om\to 0$
of~(\ref{si_1}), which with $\lim_{\om\to 0} (1-\E^{-\beta\om})/\om=\beta$ 
yields
\begin{equation}
  \re \sigma_{\rm dc} =
  \frac{\pi\beta}{ZL} 
  \sum_{n,m}^{D-1}\E^{-\beta E_n}|\langle n|\hat{\jmath}|m\rangle|^2 
  \, \delta(E_m-E_n)\,.
  \label{si_dc}
\end{equation}
$\re \sigma_{\rm dc}$ essentially counts the
number of thermally accessible current carrying (degenerate) states. 
 Since we have $\langle 0|\hat{\jmath}|0\rangle=0$ the conductivity
  almost vanishes for small $T$ below the finite size gap between
  the ground state and the first excited state.
  In the thermodynamic limit, $L\to\infty$, 
  $\re \sigma_{\rm dc}$ is related to the charge
  stiffness $D_c$, or the so-called Drude weight (at $T=0$).\cite{SS90}

The temperature dependence of the dc conductivity is illustrated in
Fig.~\ref{sigma_dc}.  Again the weak coupling results appear to have a
rather strong finite-size dependence (note that $\re \sigma_{\rm dc }$
depicted in Fig.~\ref{sigma_dc} is an intensive quantity).  When $L$
increases a pronounced peak develops at low temperatures.  This peak
can be attributed to a normal ``metallic-like'' behavior.
Independent of the coupling strength $\ep$ the polaron is an
  itinerant quasiparticle thus, for $T\to0$, always leading 
  to band conduction. When $\ep>0$, we expect that 
  $\re \sigma(\om\to 0)$ is finite
  for $T > 0$ in contrast to an ideal conductor. 

 The shoulder observed for the 10 and 12 site systems at
$T/t\gtrsim 1$, again is an artifact of our phonon 
truncation procedure as can be seen by comparing 
the data obtained for $L=6$ and
different numbers of the phonon cut-off $M$. In the strong-EP-coupling
polaronic regime, band-like transport becomes extensively suppressed
(the Drude weight is exponentially small).  Nevertheless quantum
zero-point phonon fluctuations cause polaron delocalization at $T=0$.
At higher temperatures incoherent polaron hopping transport manifests
in the temperature dependence, leading to the well-known absorption
maximum in $\re \sigma_{\rm dc}(T)$ (cf. Fig.~\ref{sigma_dc}, lower
panel). Since this signature is related to (rather local) polaron
excitation processes the position of the maximum is almost independent
of the system size. In comparison to the  
$d=\infty$ (DMFT) results~\cite{FC03} we find the same 
qualitative behavior in the relevant
temperature regime $\omega_0 \lesssim T \sim 2 \ep$ 
but a different location of the conductivity maximum.
Generally in DMFT the activation energy for polaron hopping turns out to be
lower than expected from commonly accepted arguments for finite-d systems.
Increasing the lattice size, our 1d Holstein 
data indicates that this discrepancy 
does not necessarily imply the failure of standard theory 
of hopping conduction~\cite{BB85} but may partly
arise from dimensionality effects on
polaron transport in infinite dimensions. 
Conversely, and in light of the deviations 
found for the ac conductivity (cf. Fig.~\ref{acsigma_sc_lf}),
the standard (anti-adiabatic) strong-coupling description can only be supposed 
to provide estimates on relevant energy scales in the intermediate adiabatic
EP-coupling regime. 
\section{Summary}
In this work, we have investigated the motion of a charge carrier in
response to ac and dc external fields for strongly correlated
1d electron-phonon systems.  The combination of Lanczos diagonalization
and the Kernel Polynomial Method has enabled us to calculate for the first
time quasi-exactly the temperature dependence of the optical
absorption spectra and the dc conductivity in the framework of the
one-dimensional Holstein model. Besides the well-known
polaron maximum  a pronounced absorption feature at about
$2t$ is found in the optical conductivity. 
Finite-size effects were identified and
assessed, e.g., on the basis of the $f$-sum rule.  In the physically 
most interesting range of intermediate coupling strengths and phonon
frequencies, we find that the conductivity deviates from
the standard small polaron results.

\begin{acknowledgments}
  We are grateful to M. Hohenadler and J. Loos for
  helpful discussions. This work was supported by the Deutsche
  Forschungsgemeinschaft through SPP1073, by KONWIHR and by the
  Australian Research Council. H.~F. acknowledges the hospitality at
  the University of New South Wales sponsored by the Gordon Godfrey
  Bequest. Special thanks go to NIC J\"ulich and HLRN Berlin for
  granting access to their supercomputer facilities.
\end{acknowledgments}

%\bibliography{./ref} 
%\bibliography{ref} 
\bibliographystyle{apsrev}

\end{document}